\documentclass[final,1p,times]{elsarticle} 
\usepackage{graphicx} 
\usepackage{amssymb} 
\usepackage{amsthm} 
\usepackage{lineno} 

\journal{Nuclear Physics A} 
\begin{document} 

\begin{frontmatter} 


\title{Deep Inelastic Scattering from the AdS/CFT correspondence}

\author{Anastasios Taliotis }

\address{Department of Physics, The Ohio State University, Columbus, OH 43210, USA}

\begin{abstract} 

We calculate \cite{Albacete:2008ze} the cross section of an ultra relativistic nucleus scattering on a $q\overline{q}$ pair at large coupling in $\cal N$=4 SUSY gauge theory. We study the problem in the context of the AdS/CFT correspondence \cite{Maldacena:1997re}. The nucleus is modeled as a gravitational shockwave in an $AdS_5$ background moving along the light cone. The dipole ($q\overline{q}$) is represented by a Wilson loop moving in the opposite direction. Due to the correspondence, calculating the scattering amplitude of the Wilson loop with the nucleus, reduces to calculating the extreme value of the Nambu-Goto action for an open string. Its two end points are attached to the $q\overline{q}$ respectively and it hangs in an $AdS_5$ shockwave spacetime. Six solutions are found two of which are physically meaningful. Both solutions predict that the saturation scale $Q_s$ at high enough energies becomes energy independent; in particular it behaves as $Q_s\propto A^{1/3}$ where A is the atomic number. One solution predicts pomeron intercept $\alpha_p=2$ and agrees with \cite{Brower:2006ea}. However, there is a parameter window of $r$ (dipole size) and s (c.m. energy) where it violates the black disk limit. On the other hand, the other solution respects this limit and corresponds to pomeron intercept $\alpha_p=1.5$. We conjecture that this is the right value for gauge theories at strong coupling.

\end{abstract} 

\end{frontmatter} 


The AdS/CFT correspondence provides analytical methods for dealing with problems in gauge theories when the (t'Hooft) coupling is large. It may be applied to problems related to particle collisions \cite{Albacete:2008ze}, \cite{Brower:2006ea}-\cite{Taliotis:2009} in a regime where traditional perturbative methods fail.

\begin{figure}
\centering
\includegraphics[scale=0.28,angle=0]{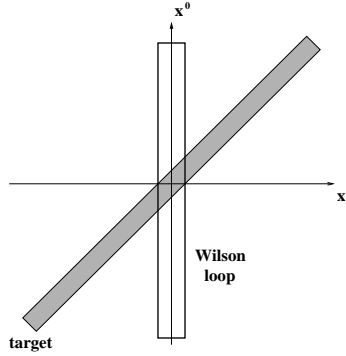}
  \caption{Dipole-nucleus scattering in the dipole's rest frame. 
    While quark and anti-quark have the same $x^3$-coordinate, we show
    them apart from each other for illustration purposes.}
  \label{rest}
\end{figure}

In this paper we investigate the DIS of a (virtual) photon that splits into a $q\overline{q}$ pair and scatters on a large nucleus (Fig. \ref{rest}). The coupling is tuned to be large and is constant throughout the interaction because the $\beta$-function of $\cal N$=4 SUSY theories is zero. We calculate the non-abelian part of the cross section which factors as
\begin{equation}\label{csec}
\sigma_{tot}^{\gamma*A}(Q^2,x_{B_{j}})/\mathrm{trans.\hspace{0.02in} area}=\int{\frac{d^2r}{2\pi}\Phi(r)N(r,Y)}\hspace{0.03in}.
\end{equation}
$\Phi$ encodes all the QED effects while $N$ is the forward scattering amplitude of the interaction. Here $Q^2\propto 1/r^{2}$ (transverse dipole size) and $Y=\mathrm{ln} \hspace{0.01in}1/x_{B_{j}}$ with $x_{B{_j}}$ the Bjorken-x. The above equation assumes the following approximations: Infinitely large nucleus in the transverse directions and therefore the impact parameter plays an overall (trivial) multiplicative role. This is precisely the meaning of the denominator on the left hand side of the (\ref{csec}). Also we suppose a homogeneous nucleus which implies the problem has no angular dependence as the dipole's orientation is irrelevant. Lastly we ignore time effects by considering a large enough extent of the nucleus along its direction of motion. This last approximation allows us to search for a static string configuration.
Our goal is to apply AdS/CFT in order to calculate the amplitude $N$. $N$ is related to the S-matrix by $N=1-S$ while the S-matrix is related to the Wilson loop formed by the $q\overline{q}$. Here is where AdS/CFT enters since it allows us to associate the Wilson loop with string theory in a rather simple way \cite{Maldacena:1998im}:       $W\propto \mathrm{exp}[i S_{NG}]$ with $S_{NG}$ the Nambu-Gotto action 

 \begin{equation}\label{SNG}
S_{NG}(r,s)=\frac{1}{2\,\pi\,\alpha'} \int d\tau \, d\sigma \,
  \sqrt{-\mbox{det} g_{\mu\nu} (X) \,\partial_{\alpha} X^{\mu}
  \, \partial_{\beta} \, X^{\nu}} \hspace*{1cm} \alpha, \beta=t,x\hspace{0.03in}.
\end{equation}

$X^{\mu}(t,x)=(t,x,0,0,z(x))$ parameterizes the static string configuration with $x\in[-\frac{r}{2},\frac{r}{2}]$ and boundary conditions $z(\pm r/2)=0$. The coordinate $z$ describes the fifth dimension of the AdS space while the other four coordinates cover the space at the boundary (Minkowski space) where the gauge theory lives. This action describes the worldsheet in a suitably chosen background gravitational field that mimics the nucleus. The nucleus is assumed to move along the $x^+$ axis as in Fig. \ref{rest}. In particular the metric $g_{\mu \nu}$ of spacetime where the string is hanging into is given by  
\begin{equation}\label{ds}
  ds^2=\frac{L^2}{z^2}\left[-2\,dx^+dx^-+\frac{\mu}{a}
      \,\theta(x^-)\,\theta(a-x^-)\,z^4\,dx^{-2}+dx_{\perp}^2+dz^2\right].
\end{equation}
This is an exact solution of Einstein's equations in vacuum with a negative cosmological constant \cite{Janik:2005zt}. $L$ is the radius of $AdS_5$. $\mu$ is proportional to the stress energy tensor $T_{ij}$ of the target (nucleus) \cite{Albacete:2008vs} and only the $T_{--}$ component is non zero. The longitudinal extent of the nucleus is denoted by $a$  and behaves as $a\propto A^{1/3}\Lambda/s$. $A$ is the atomic number, $\Lambda$ some transverse scale characterizing the system and $s=\sqrt{\mu/2a}$ is the center of mass energy \cite{Albacete:2008ze}. The equation that calculates $N$ is

\begin{equation}\label{N}
N(r,Y)=1-S(r,Y)=1- Re\left[\frac{<W>_{\mu}}{<W>_{\mu \rightarrow0}}\right]=1-Re\left[e^{i[S_{NG}(\mu)-S_{NG}(\mu \rightarrow 0)]}\right].
\end{equation}
The right hand side of (\ref{N}) is a direct application of the AdS/CFT correspondence \cite{Maldacena:1997re},  \cite{Maldacena:1998im}, \cite{Witten:1998qj}. The subtraction of the $S_{NG}(\mu \rightarrow 0)$ removes self interactions (of $q\overline{q}$). We extremize (\ref{SNG}) with the given boundary conditions $z(\pm r/2)=0$ and apply the result to (\ref{N}). The final result for $N$ turns out to be given parametrically by the set of the following two equations

\begin{equation}\label{N1}
  N (r, s) \, = \, 1 - \exp \bigg\{ i\frac{\sqrt{\lambda}\,a}{\pi\,c_0\sqrt{2}} \,
  \left[\,\frac{c_0^2\,r^2}{z_{max}^3} - \frac{2}{z_{max}} + 
    2 \sqrt{s}\,\right] \bigg\}
\end{equation}
\begin{equation}\label{zmax}
 c_0 \, r \, = \, z_{max} \, \sqrt{1-s^2\,z_{max}^4}\hspace{0.05in}.
\end{equation}

$c_0$ is a constant given by $c_0 \, \equiv \, \Gamma^2 \left(\frac{1}{4}\right)/(2 \,
    \pi)^{3/2}$ and $\lambda$ is the (large) t'Hooft coupling. We have also replaced $Y\rightarrow \mathrm{ln}\hspace{0.02in} s$ since we consider the high energy limit of the process. Finally, as can be seen from (\ref{zmax}), $z_{max}$ is explicitly a function of both $r$ and $s$; i.e $z_{max}=z_{max}(r,s)$. To see this, one has to invert (\ref{zmax}) and solve for $z_{max}$. This point is crucial as (\ref{zmax}) is a sixth degree polynomial equation for $z_{max}$ which together with (\ref{N1}) imply that the scattering amplitude $N$ is described by six different functions (branches) of r and s: each branch of the (generally complex) $z_{max}$ gives rise to a different scattering amplitude. The complex valued $z_{max}$ should be understood as a saddle point of the string theory partition function rather than a point on the (real) string. It is well known that real integrals in real variables when evaluated with the method of steepest descent, may give complex saddle points (Airy integrals provide such an example). A similar situation appears elsewhere \cite{Albacete:2008dz}.

Obviously $N$ must be given by a single function. The existence of six possible solutions comes from the fact that we have not solved the full string theory problem. Instead, we have approximated it by its saddle points which happen to be more than one. Consequently, unless we solve the complete problem, we are forced to eliminate physically meaningless solutions in a rather artificial way. Our criterion is whether they give the right asymptotics. Requiring $N(r=0,s)\rightarrow0$ (color transparency) and $N(r\rightarrow \infty,s)\rightarrow 1$ (black disk limit) two solutions are remaining. We argue that quantum corrections should in principle make the right selection. However, at this level of accuracy, we are forced to consider both of the remaining solutions as neither violates any fundamental principle.  

\begin{figure}
\centering
\includegraphics*[scale=0.239]{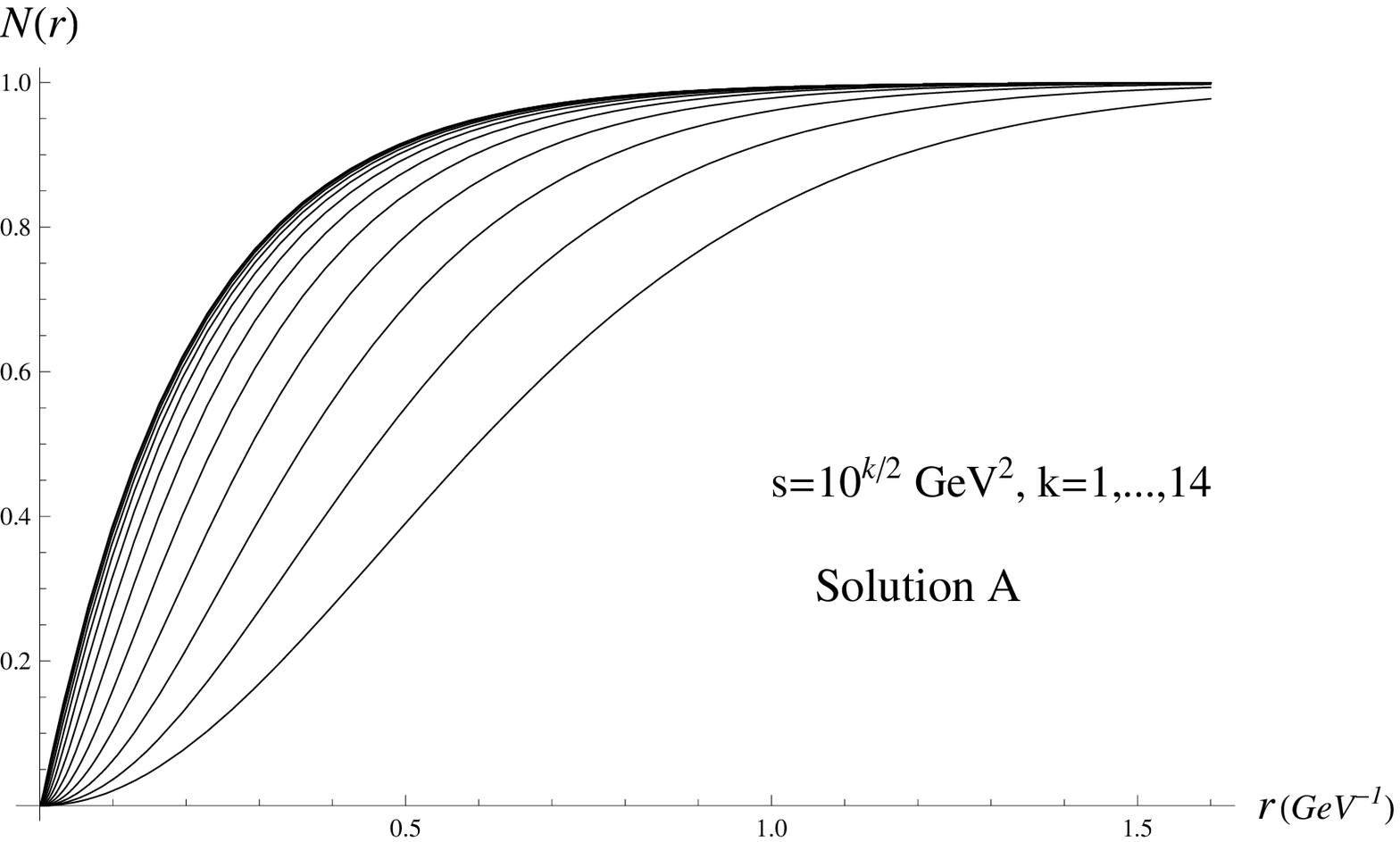}
\includegraphics*[scale=0.239]{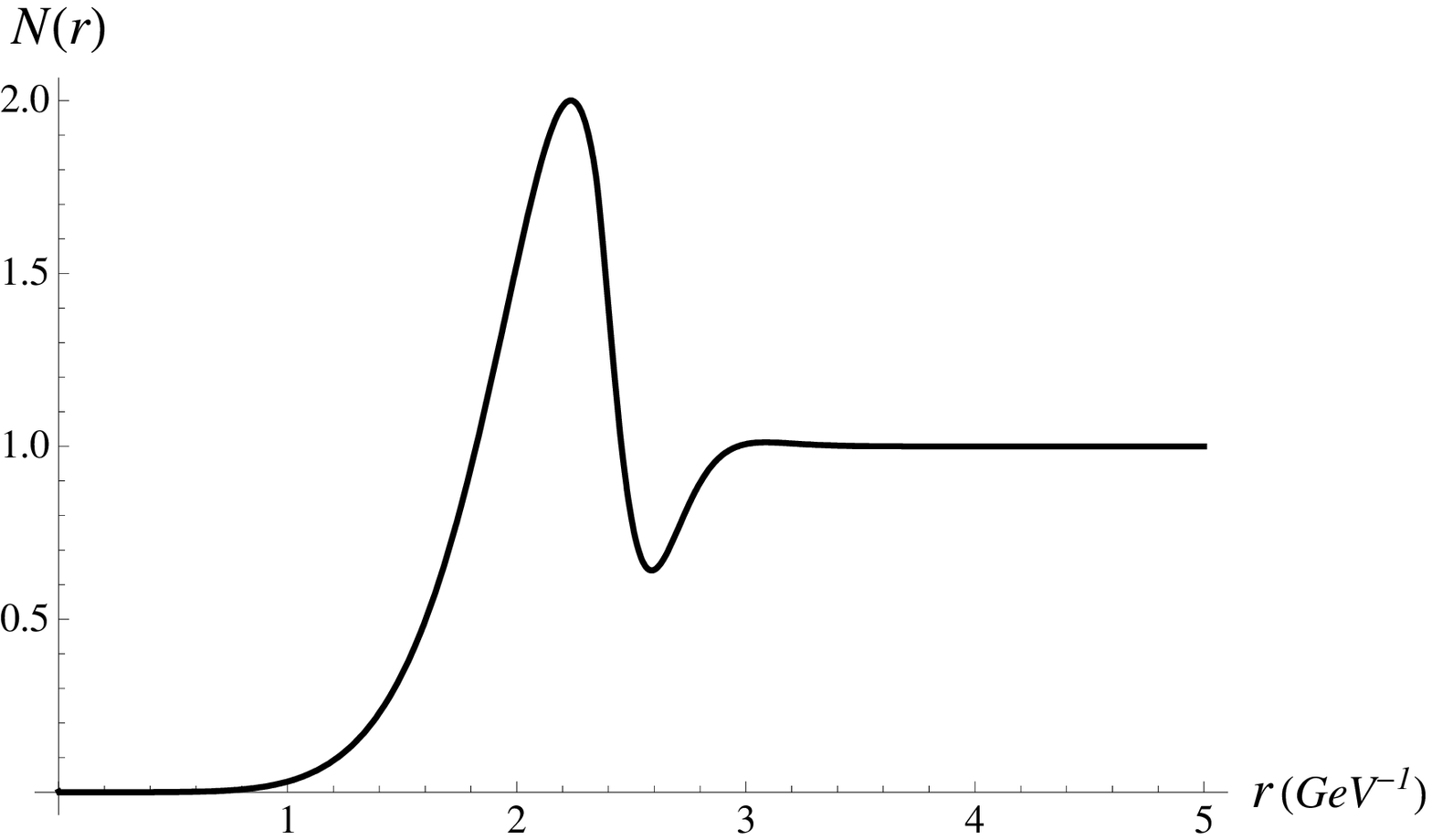}
  \caption{ {\scriptsize The Dipole scattering amplitude $N(r, s)$ (eq. (\ref{N})) for $\lambda=20$, $A^{1/3}=5$ and $\Lambda = 1$~GeV. \textbf{Left:} This is the first solution . $s$ increases from right to left. \textbf{Right:} This $N(r, s=\mathrm{fixed})$ belongs to the  second solution for some fixed $s$.}}
  \label{ndip1}
\end{figure}

The first branch gives a strictly imaginary $z_{max}$. At small-$s$ asymptotics ($r^2 s<<1$) the amplitude behaves like $N\propto \Lambda r^2 s^{1/2}$. Identifying this lower energy behavior with single pomeron exchange in the gauge theory, we find that the pomeron intercept is
\begin{equation}\label{ap1}
\alpha_P=1.5\hspace{0.03in}.
\end{equation}

In the other limit $r^2s>>1$, N becomes energy independent: $N(r,s)=N(r)$. Defining the saturation scale $Q_s$ by $N(r=1/Q_{s},s)=o(1)$ we find that the saturation scale is also energy independent and behaves like $Q_s\propto  A^{1/3}\Lambda$. A family of plots of $N=N(r,s=\mathrm{fixed})$ as a function of $r$ for several fixed values of $s$ is depicted in the left panel of Fig. \ref{ndip1}. As energy increases, the curves move to the left and in the strict infinite limit, they converge. A nice feature of this solution is that $N\rightarrow0$ as $r\rightarrow0$ and  $N\rightarrow 1$ as $r\rightarrow \infty$. The transition from small $r$ to large $r$ can be shown to be strictly monotonic for any value of $s$. The only problem with this solution is that it does not map onto Maldacena's vacuum solution \cite {Maldacena:1998im} in the limit where the gravitational perturbation (the shockwave) due to the nucleus is turned off ($\mu\rightarrow 0$).
\begin{figure}
\centering
  \includegraphics[scale=0.27]{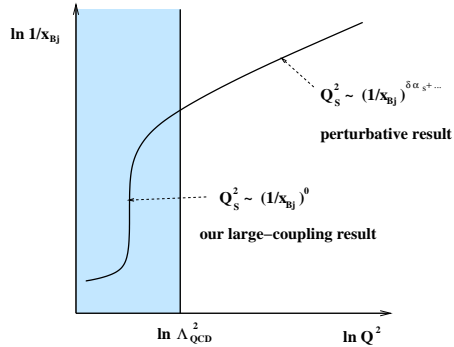}
  \caption{A sketch of the saturation line in ($\ln Q^2, \ln 1/x_{Bj}$)-plane 
    combining both perturbative (previously known) and
    non-perturbative results found in this work.}
  \label{satmap}
\end{figure}

While it is not clear whether such a matching should occur, the other solution does map to the vacuum solution in the limit $\mu\rightarrow 0$. This branch also predicts energy independent saturation and has the same limits as before at $r=0$ and $r=\infty$. At the region where  $r^2 s<<1$ we obtain $N\propto s^2$. Identifying this with a double pomeron exchange we find the intercept to be $\alpha_p=2$ in agreement with \cite{Brower:2006ea}. A feature of this branch is that the transition of the amplitude from $N=0$ to the black disk limit ($N=1$) is  not monotonic but it oscillates. This is shown in the right panel of Fig. \ref{ndip1} for a fixed value of $s$. In particular $N$ can reach as high as $N=2$ violating the black disc limit. Strictly speaking, this behavior does not violate any fundamental principle because elastic dominance allows $N$ to have range $0\leq N\leq2$. Similar conclusions are derived in \cite{Brower:2006ea}. However, these oscillations do not arise in perturbative calculations. We are thus led to the conclusion that since oscillations have no physical origin, probably this branch is not the correct one.  

Therefore our conclusion is that the pomeron intercept at large coupling is $\alpha_P=1.5$. In addition we predict that at high energies the saturation scale is energy independent and grows like $A^{1/3}$ (see Fig. \ref{satmap}). We point out that both of the physical solutions we found could be correct at this level. The only way to filter out the right one is to include quantum corrections which correspond to the functional determinants that multiply each of the two saddle points. We believe that one of the coefficients (determinants) must be suppressed compared to the other. 
\\



\textbf{Acknowledgments:} This research is sponsored in part by the U.S. Department of Energy under Grant No.
DE-FG02-05ER41377.


\begin{thebibliography}{00} 
  
 \bibitem[Albacete et~al.(2008)]{Albacete:2008ze}
J.~L. Albacete, Y.~V. Kovchegov, and A.~Taliotis, \emph{JHEP} \textbf{07}, 074
  (2008), {0806.1484}.


\bibitem{Maldacena:1997re}
J.~M. Maldacena, 
  {\em Adv. Theor. Math. Phys.} {\bf 2} (1998) 231--252,


\bibitem{Brower:2006ea}
R.~C. Brower, J.~Polchinski, M.~J. Strassler, and C.-I. Tan, 
   {\em JHEP} {\bf 12} (2007) 005,
   [hep-th/0603115]
   
\bibitem{Kovchegov:2007pq}
Y.~V. Kovchegov and A.~Taliotis, 
    {\em Phys. Rev.} {\bf C76} (2007)
  014905, 
  arXiv:0705.1234

 \bibitem[Janik and Peschanski(2006)]{Janik:2005zt}
R.~A. Janik, and R.~Peschanski, \emph{Phys. Rev.} \textbf{D73}, 045013 (2006),
  [{hep-th/0512162}].

 \bibitem{Albacete:2008vs}
J.~L. Albacete, Y.~V. Kovchegov, and A.~Taliotis, 
  {\em JHEP} {\bf 07} (2008) 100,
  arXiv:0805.2927

\bibitem{Gubser:2009}
 S.~S.~Gubser, S.~S.~Pufu and A.~Yarom,
  arXiv:0902.4062 [hep-th].

 \bibitem{Taliotis:2009}
  J.~L.~Albacete, Y.~V.~Kovchegov and A.~Taliotis,
  JHEP {\bf 0905}, 060 (2009)
  arXiv:0902.3046 [hep-th].

 \bibitem[Maldacena(1998)]{Maldacena:1998im}
J.~M. Maldacena, \emph{Phys. Rev. Lett.} \textbf{80}, 4859--4862 (1998),
  {hep-th/9803002}.


 \bibitem{Witten:1998qj}
E.~Witten, 
  {\em Adv. Theor. Math.
  Phys.} {\bf 2} (1998) 253--291,
  [hep-th/9802150]
  
\bibitem{Albacete:2008dz}
  J.~L.~Albacete, Y.~V.~Kovchegov and A.~Taliotis,
  Phys.\ Rev.\  D {\bf 78}, 115007 (2008)
  arXiv:0807.4747 [hep-th].






\end{thebibliography}
\end{document}